# Energy-Efficiency Prediction of Multithreaded Workloads on Heterogeneous Composite Cores Architectures using Machine Learning Techniques

Hossein Sayadi

Department of Electrical and Computer Engineering
George Mason University, Fairfax, VA, USA

**ABSTRACT**

Heterogeneous multi-core architectures have emerged as a promising alternative for homogeneous architectures to improve the energy-efficiency of computer systems by allowing each application to run on a core that matches resource needs more closely than a one-size-fits-all core. Composite Cores Architecture (CCA), a class of dynamic heterogeneous architectures enabling the computer system to construct the right core at run-time for each application by composing cores together to build larger core or decomposing a large core into multiple smaller cores. While this architecture provides more flexibility for the running application to find the best run-time settings to maximize energy-efficiency, due to the interdependence of various tuning parameters such as the type of the core, run-time voltage and frequency and the number of threads, it makes it more challenging for scheduling. Prior studies mainly addressed the scheduling problem in CCAs by looking at one or two of these tuning parameters. However, as we will show in this paper, it is important to concurrently optimize and fine-tune these parameters to harness the power of heterogeneity in this emerging class of architectures. In addition, most previous works on CCA mainly study traditional single threaded CPU applications. In this work, we investigate the scheduling challenges for multithreaded applications for CCA architecture. This paper describes a systematic approach to predict the right configurations for running multithreaded workloads on the composite cores architecture. It achieves this by developing a machine learning-based approach to predict core type, voltage and frequency setting to maximize the energy-efficiency. Our predictor learns offline from an extensive set of training multithreaded workloads. It is then applied to predict the optimal processor configuration at run-time by taking into account the multithreaded application characteristics and the optimization objective. For this purpose, five well-known machine learning models are implemented for energy-efficiency optimization and precisely compared in terms of accuracy and hardware overhead to guide the scheduling decisions in a CCA. The results show that while complex machine learning models such as MultiLayerPerceptron are achieving higher accuracy, after evaluating their implementation overheads, they perform worst in terms of power, accuracy/area and latency as compared to simpler but slightly less accurate regression-based and tree-based ML classifiers.

**Keywords-** Heterogeneous architectures, Composite cores, Machine learning, Scheduling, Multithreaded applications, Energy-efficiency

## 1. INTRODUCTION

Heterogeneous multi-cores offer an effective solution to energy-efficient computing. Heterogeneous architectures integrate multiple cores with various flavors on the same die, where each core is tuned for a certain class of workloads and optimization goals (either performance or power consumption). To unlock the potential of the heterogeneous design, software applications must adapt to the variety of different processors and make good use of the underlying hardware by executing workloads on the most appropriate core type. By running multithreaded applications on heterogeneous architectures, each thread is able to run on a core that matches its resource requirements more closely than one size fits all solution [1]. Commercially available heterogeneous architectures include Intel Quick IA [8], ARM's big.LITTLE [10], AMD Fusion APUs [50], and Nvidia Tegra 3 [7] that integrates high performance big cores with low power little cores on a single chip.

Although heterogeneous architectures take advantage of application characteristic variation at run-time and improve energy-efficiency, they create unique challenges in effective mapping of threads to cores. As the core configurations in heterogeneous multicores become more divers, they become more difficult to program effectively. In other words, the effectiveness of heterogeneous architectures significantly depends on the mapping and scheduling policy and how efficiently we can allocate applications to the most appropriate processing core [3, 5, 11]. Applying ineffective scheduling decisions can lead to performance degradation and excess power consumption in such architecture.

Composite Cores Architectures can provide further benefits by allowing the system to construct a right core for each running application. Several designs have been proposed that provide some level of dynamic heterogeneity [1, 14, 17, 21, 22]. In [14] and [21] the concept of composite cores architecture is proposed where a big core architecture can dynamically decomposed intro a smaller little-core architecture. The authors in [1] adapted the concept of composite cores in 3D by further enabling the core composition and decomposition at a low granularity of processor building blocks such as register file and load and store queue. Their proposed architecture allows multiple smaller cores to be composed together to build a larger core or vice versa, as needed. While CCA provides more opportunity to construct the right core for the running applications, it is making the scheduling a difficult problem.

Previous studies have mainly examined the advantages of using single threaded applications in CCA. However, running multithreaded applications on CCA and composing ideal processor architecture for energy-efficiency is a more challenging problem, considering the possible number of cores and threads, type of core micro-architecture, or combinations of core types. Furthermore, the challenge of how many and what type of core to compose for each multithreaded application becomes even more complicated considering the impact of other tuning parameters on energy-efficiency such as operating voltage and frequency. In this work, we focus on the benefits of running multithreaded applications on CCA and how this architecture provides opportunities to improve the energy-efficiency.

The main challenge for scheduling is to effectively tune system, architecture and application level parameters in CCA when running multithreaded applications. The particular parameters that are critical to performance and power considered in this work include core type, voltage/frequency settings and the number of running threads. While there has been a number of works on mapping applications to heterogeneous architectures, no solution has been developed for mapping multithreaded applications into CCA with its unique architecture. In addition, previous studies on mapping applications to multi-core architectures have focused primarily on 1) homogeneous architectures, 2) static heterogeneous architectures where the number and type of cores are fixed at design time, and 3) configuring individual or a subgroup of tuning parameters at a time, such as application's thread counts [7, 11, 26, 27], voltage/frequency [2,3], or core type [6, 10, 14, 21, 22, 23, 27] and they have ignored the interplay among all of these parameters. This study indicates that these parameters individually, while important, do not make a truly optimum configuration to achieve the best energy-efficiency on a CCA. The best configuration for a multithreaded application can be effectively found, only when these parameters are jointly optimized.

In this paper, through methodical investigation of power and performance, and comprehensive system and micro-architectural level analysis, we first characterize multithreaded applications on CCA to understand the power and performance trade-offs offered by various configuration parameters and to find how the interplay of these parameters affects the energy-efficiency. Our study is focused on a CCA where many little cores (base) can be configured into few big cores (composed) and vice versa. The experimental results support that there is no unique solution for the best configuration for different applications. Given the dispersed pattern of optimum configuration, we have developed various Machine Learning (ML) models to predict energy-efficiency, and guide scheduling and fine-tuning parameters to maximize the energy-efficiency. As behavior of applications changes at run-time, we applied our prediction and tuning method at a fine-grained level of individual parallel region within an application.

We used five well-known machine learning models to build predictors based on the knowledge extracted from an extensive set of hardware performance data which are a good representative of application behavior for each parallel region within the training phase. The models are then used at run-time to predict the optimal processor configuration for each parallel region of a multithreaded workload to maximize the energy-efficiency.

This paper in brief makes the following contributions:

- Through methodical investigation of power and performance results, we characterize parallel regions of various multithreaded applications on a CCA and demonstrate how the interplay among various application, system, and architecture level parameters affect the performance and energy-efficiency of those parallel regions.
- We develop various machine learning-based models to predict the energy-efficiency of parallel regions for various configurations of CCA for a wide range of application, system and architecture level parameters.
- We analyze the proposed machine learning-based models in terms of their prediction accuracy, power overhead and implementation overheads to understand their cost effectiveness.

The remainder of this paper is organized as follows. Section 2 describes previous studies on scheduling challenges in heterogenous architectures and motivation of this work. Section 3 presents an overview of our proposed approach. The experimental setup details are given in Section 4. Section 5 presents the characterization results and provides the performance and energy-efficiency analysis of multithreaded applications on CCA. Next, the proposed machine learning-based approach for energy-efficiency prediction and scheduling in CCA is explained and evaluated in detail in section 6. Finally, Section 7 presents the conclusion of this study.

## 2. BACKGROUND AND MOTIVATION
### 2.1 Heterogeneous Architectures

Static heterogeneous architectures have existed in many forms, including Intel Quick IA [8], ARM's big.LITTLE [10], TI OMAP 5 [41] and Nvidia Tegra 3 [9] which integrates a high performance big core with low power little core on a single chip. The static heterogeneous architecture enables efficient thread-to-core mapping and permits a change in the mapping across phases of execution through thread migration.

Prior research has shown that the potential benefit of a static heterogeneous architecture is greater with fine-grained thread migration than with coarse-grain migration [14]. In [12] an Intel Xeon is integrated with an Atom processor. Code instrumentation is used at the function or loop level to schedule different phases of the application on each processor. However, the separate core and memory subsystems in static heterogeneous architectures incur power and performance overheads for application migration, which makes dynamic mapping ineffective for fine-grained migration.

Unlike static heterogeneous architecture where the number and type of cores are fixed at run-time, dynamic heterogeneous architectures can be configured at run-time [1, 5]. This provides more opportunity to map an application to a core which matches its resource needs more closely [3, 4]. Some of the first efforts to provide this kind of heterogeneity include Core Fusion [21] and TFlex [22]. Composite core proposed a dynamic heterogeneous architecture where a big core can dynamically be decomposed into a smaller little-core.

The work in [1] and [2] extended the concept of composite core into 3D stacking which enables very fine-grain sharing of resources between cores on a stacked chip multiprocessor architecture. Their proposed architecture allows multiple smaller cores to be composed together to build a larger core or vice versa, as needed. Previous work on dynamic heterogeneous architecture in general and composite core in particular have mainly studied mapping of single threaded applications. This work is different as it mainly focuses on multithreaded applications and how they would benefit from such architecture to maximize the energy-efficiency. Similar to [1, 2, 3, 4], we are assuming that big cores (composed) are constructed by composing multiple little (base) cores.

### 2.2 Scheduling Challenges in CCA

As mentioned before, a main challenge for heterogeneous architectures is the mapping and scheduling decision, which finds the most efficient application-to-core match at run-time. The researches in [11] and [29] address the problem of dynamic thread mapping in static heterogeneous many-core systems. Prior research aimed to maximize performance under power constraints. Our work is different as it first targets dynamic heterogeneous architectures where core size can be adapted at run-time, and second it aims to maximize the energy-efficiency by reducing the energy-delay. It is important to note that the power and performance of an application on different cores at various frequencies must be known for proper mapping. Traditional designs suggest selecting the best core based on a small sampling of applications on each core [26]. Other techniques [5, 6, 14], estimate core performance without running applications on a particular core type. The work in [6] and [14] provide a model for performance estimation on two core types (i.e., big and little cores). The complexity of application mapping on a heterogeneous architecture increases exponentially with an increasing number of core types and applications [33].

There have been several works on mapping multithreaded applications on homogeneous architectures. The work in [7] suggested a framework called "Thread Reinforcer" to determine the appropriate number of threads for a multithreaded application on a homogeneous architecture. It examines the mapping between number of threads and number of cores to find the optimal or near optimal number of threads to minimize the execution time. The research in [6] proposed a scheduling method to predict application to core mappings that enhances performance. Using profiling parameters, it estimates performance and examines whether the workload needs to run on different core type. The work in [11] proposed a mapping strategy for multithreaded applications on static heterogeneous multicore architecture by initializing a maximum throughput mapping and iteratively performing a thread swap on adjacent types of cores until the power constraint is met. The research in [5] took a closer look at joint optimization of voltage and frequency as well as the microarchitecture. It proposed a platform, which is capable of scaling resources, i.e., bandwidth, capacity, voltage, and frequency, based on single-threaded application performance requirements at run-time while reducing EDP.

### 2.3 Motivation of this Study

Multithreaded applications are composed of a number of parallel regions which are separated by serial regions. In this work, we refer to these parallel regions as Region of Interest (ROI). In a homogenous multicore architecture and for conventional scheduling, all of these regions are processed on the same core type, same voltage and frequency, and number of threads, though not all regions may have the same preferences. Some of these ROIs may benefit from different configuration than the others to obtain the maximized energy-efficiency. As mentioned earlier, the main challenge for scheduling is to effectively tune system, architecture and application level parameters in CCA for the entire application as well as the intermediate parallel regions in order to achieve maximum energy-efficiency. In this work, similar to [1, 5], we are assuming that big cores (composed) are constructed by composing multiple little (base) cores.

Fig. 1 illustrates an overview of optimal configurations for a multithreaded application, *lu.cont*, selected from SPLASH2 benchmark suite with three parallel regions. In each ROI the best possible configuration for core type, operating frequency and thread counts that results in maximized energy-efficiency is specified. As can be seen, ROI1 needs to be executed on the base core with 2.4GHz frequency and 8 threads, whereas for ROI2 in order to achieve the best energy-efficiency, we need to compose two small cores to make a big core. Moreover, the optimal operating frequency increases to 2.8GHz. The ability to accurately predict the optimal application, system and even microarchitecture parameters and adapt them to achieve the maximum energy-efficiency within different parallel regions of an application is the main motivation for this work. We develop several machine learning-based approaches which are able to predict the optimal setting of tuning parameters and change them to suit the specific requirement of each parallel region within the application.

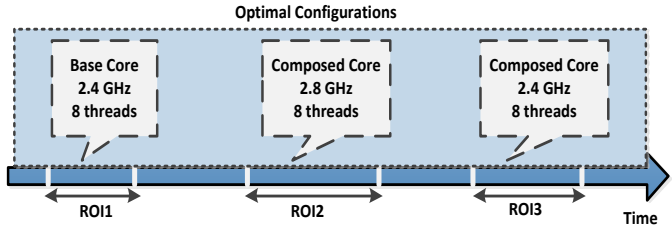

**Fig. 1. Optimal configurations (Core type, Frequency, Thread count) in different parallel regions of an application**

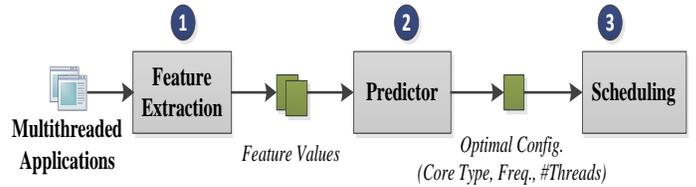

**Fig. 2. An overview of our approach for predicting the optimal configuration and scheduling the multithreaded application**

## 3. OVERVIEW OF OUR APPROACH

Fig. 2 depicts our three-stage approach for predicting the right core type and application configuration when running a multithreaded application on a composite core architecture. Our machine learning-based approach begins from extracting microarchitectural data (referred as feature extraction), from different parallel region of application to characterize the multithreaded workload. This data (or features) includes the Hardware Performance Counter (HPCs) information, which are representative of application behavior at run-time. It is notable that HPCs are a set of special-purpose registers embedded in the processing units of today's microprocessors to capture the trace of hardware events for a running application on the computer systems [12, 51, 52]. Recently, HPC information have been widely used for architectural analysis in various research domains including predicting the performance and energy-efficiency of computer systems [3, 4, 18, 40, 46], improving the security of systems such as embedded systems [12, 42, 43], and architectural analysis of various applications on server class architectures [28, 35, 38, 44] with the aid of machine learning and data mining techniques.

At next step of our proposed approach, a machine learning-based predictor (that is built off-line) takes in these features data and predicts the best configuration settings for a given parallel region. For this purpose, we have implemented five well-known machine learning algorithms and compare them in terms of accuracy and power and area overhead to find to the most effective learning model which yields in optimized energy-efficiency. Finally, we configure the processors and schedule the application to run on the predicted configuration. In this work the metric that we use to characterize energy-efficiency is the Energy Delay Product (EDP) which aims to balance performance and power consumption. We construct and compare five ML-based predictors using the machine learning algorithms described in the section 6 to guide the scheduling decisions in a CCA.

## 4. EXPERIMENTAL SETUP

This section provides the details of our experimental setup. We use Sniper [16] version 6.1, a parallel, high speed and cycle-accurate x86 simulator for multicore systems. McPAT [15] is integrated with Sniper and was used to obtain power consumption results. We study SPLASH-2 [17] and PARSEC [53] multithreaded benchmark suite for simulation. For architectural simulation, we modeled a heterogeneous composite core architecture based on the recently proposed work in [21, 22]. For base core (little) architecture, we model a core similar to the Atom Silvermont [15] and for composed core (big) we configure core resources similar to Xeon 5500 Series known as Gainestown [20]. We use the Uncore event set of Silvermont and the Intelligent Performance Counter of Gainestown to collect data for characterization and drive the scheduling and mapping algorithms. These performance counters are already in place in these architectures and we use them to extract and evaluate the actual behavior of applications (I/O, CPU or memory intensive) [25, 32, 46] for predicting the energy-efficiency and assist in scheduling decision. It is important to note that for benchmark simulation we applied the binding (one-thread-per-core) model with #threads == #cores in order to maximize the performance of multithreaded applications [3, 4, 7].

## 5. CHARACTERIZATION RESULTS

In this section, we evaluate the application performance and energy-efficiency sensitivity to the tuning parameters of operating frequency, number of running threads, and the choice of microarchitectures (base vs. composed) in heterogeneous composite cores architecture. The studied parameters not only directly impact the power and performance of the processor, but they also influence one another. For instance, as we will show application sensitivity to the number of threads varies significantly across core microarchitectures. Also, the interplay of frequency and core size affects application power and performance sensitivity. The optimal system and microarchitecture configuration to maximize energy-efficiency varies based on the characteristics of the application, which all together influence the best tuning strategy. Therefore, it is essential to investigate the interplay of these parameters to guide the optimal mapping and scheduling decision in CCA. These observations form the basis for developing the energy-efficiency prediction models presented in section 6.

Note that the entire set of benchmark analysis results is quite extensive. Therefore, due to space limitation we only present the results for a limited number of representative benchmarks shown in Fig. 3. This figure depicts the overall performance in terms of execution time (represented as a bar graph) and EDP results (represented as a line graph) for four different multithreaded benchmarks across different core types, frequencies and number of threads. In this section, first we discuss the impact of changing each parameter on energy-efficiency and next we perform a joint analysis to investigate the interplay of these parameters and their influence on energy-efficiency in heterogeneous CCA.

### 5.1 Frequency Sensitivity

For this analysis, all benchmarks were simulated using a baseline composed core running with only a single thread. The operating frequency is swept from 1.6 GHz to 2.8GHz with a step of 400MHz and the voltage is changed between 0.7, 0.8, 0.9, and 1V, respectively. As can be seen in Fig. 3, some benchmarks are very sensitive to changing the frequency. For instance, in fmm and cholesky reducing the frequency almost linearly reduces the overall performance. Overall, as expected, as the frequency increases, the performance increases accordingly. Similarly, higher frequencies use more power. The EDP results show that a higher frequency leads to a higher EDP.

Increasing the number of threads interestingly reduces the sensitivity to frequency. In other words, increasing the number of running threads increases the performance gain due to parallelization. Consequently, the overall performance as the number of threads increases is more influenced by the speedup gain as a result of parallelism rather than operating at higher frequency. Moreover, the results show that the base core is more sensitive to frequency scaling than the composed cores. This is also an interesting observation as the composed core has a large pipeline, allowing it to tolerate performance cost due to alterations in access latency to cache subsystem as a result of frequency scaling. Note that changing clock frequency changes the number of cycles it takes for the processor to communicate with the cache.

### 5.2 Core Type Sensitivity

In this section, the results are reported for a baseline configuration with a core running a single thread at the highest frequency of 2.8 GHz and operating voltage of 1V. The changing parameter is the core type, which varies between a base (little) core and a composed (big) core architecture. Core type demonstrates some variation with regards to EDP. As shown in Fig. 5, there is a clear gap between the big composed and little base cores (in Thread1 and F2.8), with composed core having lower EDP. In these cases, the performance benefits of the composed core outweigh the energy savings of the base core.

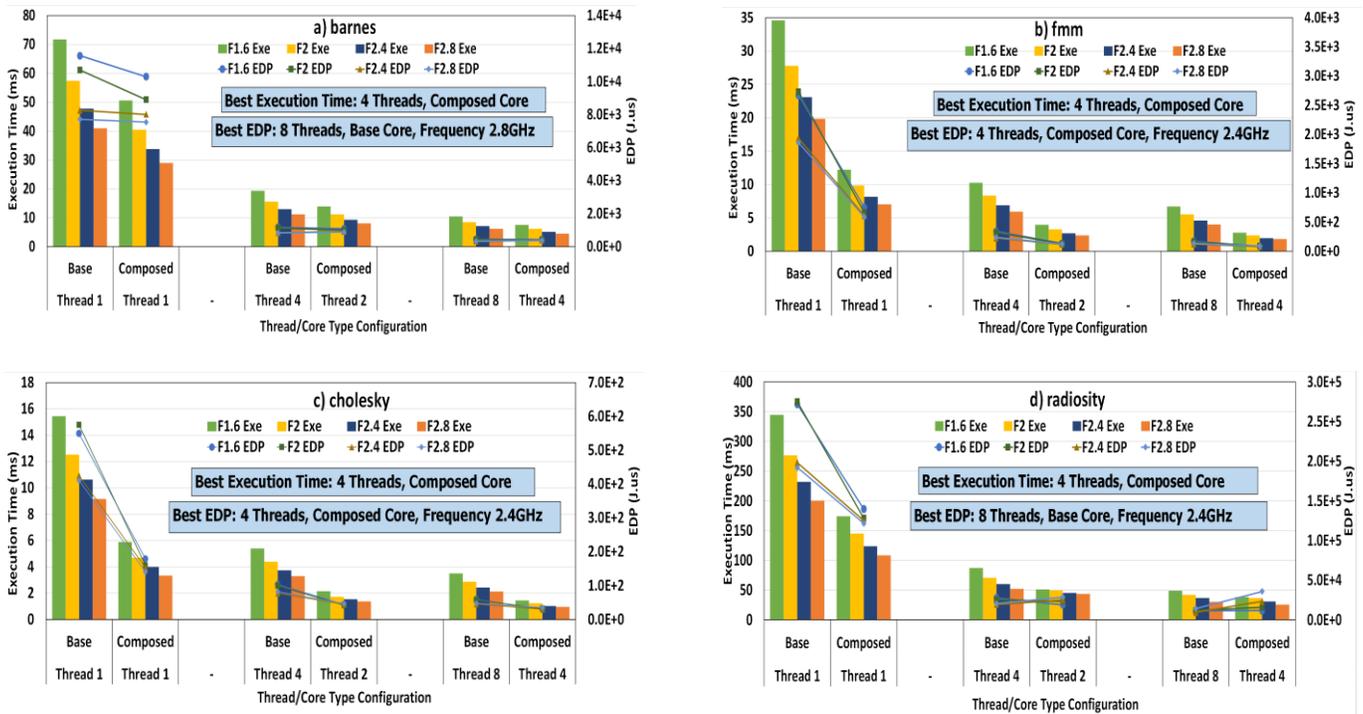

**Fig. 3. Execution Time and EDP of a) barnes, b) fmm, c) cholesky, d) radiosity with various Core Types, Threads, Frequencies**

### 5.3 Thread Count Sensitivity

Finally, each benchmark is simulated with varying the numbers of threads. In this step, each simulation was performed at the same frequency of 2.8 GHz and operating voltage of 1V, when changing the number of threads from 1 to 8. As shown in Fig. 3, increasing the thread counts leads to better performance at the cost of higher power. Moreover, there is a large gap between the EDP values of base and composed core in lower number of threads. In particular, we observe that by increasing the application thread counts, the corresponding gap between different core types diminishes and makes the base core competitive to the composed core in terms of EDP.

### 5.4 Joint Analysis of Core Type, Frequency, and Thread Count

To understand the interplay among various tuning parameters and find the optimum configuration for maximizing the energy-efficiency, in this section all permutations of the parameters were simulated. We test four voltage/frequency settings on two core types and execute each multithreaded benchmark with 1 to 8 threads using a native input set, where each thread is assigned to a single core. This yields a total of 64 possible settings corresponding to each core type, voltage/frequency and thread count combination, which are illustrated in Fig 3. Due to space limitation, we only demonstrate the results for 1, 4 and 8 running threads. As shown, the best evaluated execution time and EDP for each application are shown in each figure.

In this paper, we investigate two baseline heterogeneous CCA which consist of multiple base and composed cores: 1) 8base/4comp, and 2) 4base/2comp. Table 1 presents the optimal set of results for both architectures. This table includes benchmarks name, followed by the best core configuration parameters (Core, Freq., #Thread) in terms of EDP across base and composed cores. We have also calculated the relative EDP variation for each benchmark, which indicates the relative difference between energy-efficiency for the best configuration parameters in base and composed cores. We quantify variation as (best_base – best_comp)/best_base. The variation parameter indicates whether it is justified to compose cores. For this purpose, a variation threshold is defined that decides what type of core architecture should be selected for executing the corresponding multithreaded application more energy-efficiently. This user-defined threshold can be adjusted based on the architecture and available resources as well as the cost of core composition. Note that composing base core to build big composed cores is not free and comes with power as well as core utilization overhead. The core utilization overhead is in fact due to using additional cores to build bigger cores. When cores are composed to build a bigger core, fewer cores will be available for incoming or co-scheduled applications. In this work we assume a 20% variation threshold. As a result, if the EDP variation between best-base and best-composed architectures is found to be lower than 20%, we use the base core for scheduling instead of composing to avoid power as well as core utilization costs.

As can be seen from Table 1, for most studied applications the best running thread counts remain unchanged across different core types. For instance, *barnes* performs with 2.4 GHz and 2.8 GHz on base and composed cores, respectively, while the best number of running threads on both architectures is 8. As shown, the variation has negative value for some cases, which indicates it is more energy as well as core-utilization efficient to run the application on little base core. Therefore, given that the variation value is lower than pre-defined threshold, rather than running the application on costly big composed core, we schedule the multithreaded application onto cost-effective little base core. From these observations, we conclude that while we can obtain significant performance gains, power and core utilization costs could be drastic when running application on big composed core. As a result, in those cases we choose the little base core as the optimal core configuration.

### 5.5 Parallel Region Analysis

As explained before, multithreaded applications are composed of a number of parallel sub-regions, which are separated by serial regions. Considering the application behavior, not all ROIs may have the same performance and power requirements. To illustrate the improvements offered by composite core architectures, studied multithreaded benchmarks were modified to monitor the behavior of each parallel region within the application. Simulation markers were placed at different sections of the benchmarks that would be simulated as individual parallel regions. All simulations were then run again using all permutations of the configuration parameters discussed earlier, including voltage/frequency, core type, and thread count. Due to space limitation in our paper, we only present simulations results from parallel regions of four benchmarks which are reported in Table 2. Each table shows, for a given benchmark, the optimal configuration in terms of EDP for each of the ROIs listed. It can be clearly seen that every ROI within the application does not benefit from the same configuration parameters.

**Table 1. Optimal configurations with optimization target of EDP for different architectures**

| Benchmark | 8Base/4Comp | | | | | 4Base/2Comp | | | | |
|---|---|---|---|---|---|---|---|---|---|---|
| | Best-base | | Best-composed | | Var. (%) | Best-base | | Best-composed | | Var. (%) |
| | *Freq. (GHz)* | *#Thread* | *Freq. (GHz)* | *#Thread* | | *Freq. (GHz)* | *#Thread* | *Freq. (GHz)* | *#Thread* | |
| barnes | 2.4 | 8 | 2.8 | 4 | -444.8 | 2.8 | 4 | 2.8 | 2 | -475.4 |
| fmm | 2.4 | 8 | 2.4 | 4 | 2.2 | 2.4 | 4 | 2.8 | 2 | -2.9 |
| cholesky | 2.4 | 8 | 2 | 4 | 28 | 2.4 | 4 | 2.8 | 2 | 5.8 |
| radix | 2.8 | 8 | 2.8 | 4 | -138.7 | 2.8 | 4 | 2.8 | 2 | -236.2 |
| radiosity | 2.4 | 8 | 1.6 | 4 | -102.8 | 2.4 | 4 | 1.6 | 2 | -128.1 |
| raytrace | 2.4 | 5 | 2 | 4 | -28.9 | 2.4 | 4 | 2.8 | 2 | -152 |
| fft | 2 | 4 | 2 | 2 | 36.3 | 2 | 4 | 2 | 2 | 36.3 |
| lu.cont | 2.4 | 8 | 2.8 | 4 | 27.2 | 2 | 4 | 2 | 2 | 7.8 |
| blackscholes | 2.8 | 4 | 2.4 | 4 | 83.85 | 2.8 | 4 | 2.4 | 2 | 77.08 |
| bodytrack | 2 | 3 | 2 | 3 | 41.23 | 2 | 3 | 2 | 2 | 34.1 |
| ferret | 2 | 6 | 2 | 4 | 62.4 | 2 | 4 | 2 | 2 | 54.3 |

In order to clarify this point, here we look at an example, the radix benchmark, in more depth. Radix benchmark was instrumented with four individual sub-regions. As results show, all four sub-regions have different set of configuration parameters to achieve the best EDP. The core type varies between base and composed, and the frequency varies between 2.8GHz and 2.4GHz. Also, the thread counts changes between 7 and 8. This example demonstrates the importance of using right tuning parameters for best EDP, not only for the entire multithreaded application, but also even for each parallel region within the application. Overall, the results show the importance of concurrent optimization at the application, system and microarchitecture levels at coarse-gained level of application or even fine-grained level of individual parallel regions within the application to maximize the energy-efficiency. The challenge is to develop a technique that automatically determines the best configuration for any given multithreaded applications and optimization goal and perform the tuning at a fine-grained level of individual parallel region within the application. In the next section, we will describe our proposed approach using various machine learning-based solutions.

The diversity of optimum configurations across various applications and their parallel ROIs demonstrates that when running a given multithreaded workload on a heterogeneous CCA, depending on the application and energy-efficiency optimization metric, different configuration parameters (Core Type, V/Freq., #Thread) lead to the best energy-efficiency. The configuration also changes at run-time for each parallel region within the application. In other words, experimental results support that there is no unique solution as the best configuration across various parallel regions of an application. This dispersed pattern of optimum results implies the necessity of developing a prediction method to guide scheduling decision of multithreaded applications onto heterogeneous composite cores architecture in order to enhance the energy-efficiency.

## 6. PREDICTIVE MODELING

Methods of machine learning which build predictive models that generalize training data have proven to be effective in predicting the characteristics and behavior of applications running on computing systems [3, 4, 12, 34, 52]. Recent works have proposed linear regression modeling [24] to estimate the power and performance of a processor at run-time [3, 5, 40, 46]. As mentioned earlier, in this work we implement different machine learning models to estimate the EDP. Table 3 shows five machine learning models that we use for predicting the best processor and application configuration to deliver the lowest EDP. These machine learning-based models include least square median, linear regression, Multi-layer Perceptron (an artificial neural network model), and two decision tree techniques namely REPTree and M5Tree. We selected these five classifiers for two reasons. First, they are from three different types of machine learning methods; regression, neural network, and decision tree, covering a diverse range of learning algorithms which are inclusive to model both linear and non-linear problems. Second, the prediction model produced by these learning algorithms is deterministic which is compatible with our numerical target variable, EDP. All of ML classifiers were implemented using WEKA machine learning toolkit [36]. The inputs to our models are a set of features extracted from application profiling at run-time. While these microarchitectural features are accessible through our simulator infrastructure, in a real hardware, they are also accessible through hardware performance counters. The output of our models is the optimal EDP that corresponds to the type of core to use for running the application, the clock frequencies of the core and thread counts. The goal is to use these models to find the best configuration parameters for individual parallel region within an application and understand how these parameters should be adapted at run-time in a heterogeneous architecture.

**Table 2. Optimal configuration in different parallel regions of applications for EDP optimization**

| radix | | | |
|---|---|---|---|
| Region | Core Type | Freq. | #Threads |
| 1 | base | 2.8 | 8 |
| 2 | comp | 2.4 | 8 |
| 3 | base | 2.8 | 8 |
| 4 | comp | 2.8 | 7 |
| cholesky | | | |
| Region | Core Type | Freq. | #Threads |
| 1 | base | 2.8 | 1 |
| 2 | base | 2.4 | 8 |
| 3 | comp | 2.8 | 8 |

| fft | | | |
|---|---|---|---|
| Region | Core Type | Freq. | #Threads |
| 1 | base | 2.8 | 8 |
| 2 | comp | 2.8 | 8 |
| 3 | comp | 2.8 | 1 |
| 4 | comp | 2.4 | 8 |
| 5 | comp | 2.8 | 8 |
| lu.cont | | | |
| Region | Core Type | Freq. | #Threads |
| 1 | base | 2.4 | 8 |
| 2 | comp | 2.8 | 8 |
| 3 | comp | 2.4 | 8 |

Table 3. Machine learning classifiers used for prediction

| ML Classifier | Learning Type |
|---|---|
| LinearReg | Regression |
| LeastSqMed | Regression |
| MultiLayerPercep | Neural Network |
| M5Tree | Decision Tree |
| REPTree | Decision Tree |

Developing and deploying these models include a 3-step process for supervised machine learning as follows: (i) generating training data, (ii) developing a predictive model, (iii) using the predictor at run-time for every parallel region of the application.

### 6.1 Training the Predictor

Fig. 4 depicts the process of using training multithreaded applications to build a machine learning classifier to predict energy-efficiency. Training involves finding the best processor and application configuration and extracting feature values for each training workload, reducing the extracted features to the most vital performance counters, and developing a learning model from the training data. It is important to note that the input variables in our classifiers are extracted performance counter information from different parallel regions of application, and the output variable is the EDP for a given set of tuning parameters.

**Generating Training Data.** To derive the prediction model for energy-efficiency, we need to develop a data set to train the prediction model. We applied our ML classifiers on extensive set of SPLASH2 multithreaded benchmark suit. The studied multithreaded applications represent diverse compute, memory and I/O intensity behavior. We exhaustively execute each training benchmark, with different processor and application configurations and record the configuration with lowest EDP. We have also collected the hardware performance data from 20 parallel sub-regions of these applications and use them to build regression, neural network and decision tree classifiers for predicting the EDP. As mentioned earlier, the extracted features from each of the parallel ROIs appropriately represent the application behavior during these execution phases. For each parallel region within an application, we collect twelve HPC data on all possible configurations of core types, voltage/frequency operating points, and number of threads. We applied machine learning models on the studied benchmarks using these performance counters and profiling information for predicting energy-efficiency. The architectural information used for differentiating workloads behavior is listed in Table 4. In order to validate each of our classifiers, we applied the percentage split method to divide the dataset into two sets, using 70% (known applications) of the data to train the model and 30% (unknown applications) to simulate and test.

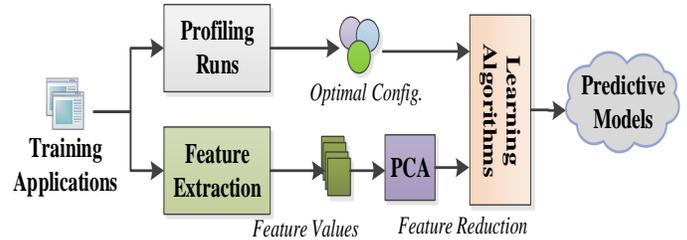

Fig. 4. Training process for machine learning predictive models

Table 4. Hardware performance data used for training the classifiers

| Category | Hardware performance counter |
|---|---|
| Memory subsystem | L1 D-cache access, L1 D-cache miss, L1 I-cache access, L1I- cache miss, L2 cache access, L2 cache miss, I-TLB miss, D-TLB miss |
| Instructions | Integer instruction issue, Integer floating point issue |
| Branch | Branch instruction, Branch misprediction |

**Developing Predictive Models.** The features together with the processor configuration are supplied to each supervised learning algorithm. The learning algorithm attempts to find a correlation from the feature values to the optimal configuration and predicts the energy-efficiency that corresponds to each configuration. As described in previous section, our predictors are based on a number of features extracted from the hardware performance counter attributes. One of the key aspects in building an accurate predictor is finding the right features to characterize the input data. We started from twelve performance counters that can be collected from parallel regions of each running multithreaded application. These features listed in Table 4 include performance counters representing pipeline front-end, pipeline back-end, cache subsystem, and main memory behaviors and are influential in the performance of standard applications. As shown in Fig. 4, after feature extraction we use Principle Component Analysis (PCA) and correlation analysis on our training set to monitor the most vital micro-architecture parameters to capture application characteristics. By applying the attribute reduction method, we determine the four most related performance counters including include L1 D-cache access, L2 cache-access, L2 cache-miss and branch misprediction. These performance counters are included in our model as input parameters.

## 6.2 Prediction Phase

Once we build our ML predictor, we deploy it for predicting the energy-efficiency of various configurations. Fig. 5 provides an overview of EDP prediction process and tuning processor and application parameters using the trained machine learning classifiers. This prediction model predicts continues values representing energy delay product as a function of performance counter inputs and tuning parameters, which is then used to make the scheduling decisions at run-time. In particular, in this phase we run a multithreaded application with the most aggressive configuration setting where all tuning parameters are set at max (maximum number of threads, highest frequency, and for composed core). It is important to note that this would be the fastest way to collect run-time features of an application, since this is done for most aggressive configuration, which corresponds to the highest performance. At run-time, we extract the hardware performance counters by profiling the application for each parallel region. The ML classifier then takes the key performance counter features and configuration settings as inputs, and outputs the system energy-efficiency for each configuration. The configuration corresponding to the lowest EDP is then used to tune and schedule the application. Thus, at run-time, given an unknown application, the predictor can predict the EDP of all possible configurations based on a single run data. The configuration corresponding to the lowest estimated EDP is then selected for the run. The predictive models by observing run-time behavior of a multithreaded application running with a specific configuration, predicts the right configuration parameters to achieve the maximum energy-efficiency. It is important to note that each predictor can be simply trained for other objective such as $ED^2P$ optimization.

## 6.3 Experimental Results

In this section, we present the evaluation results for our machine learning predictors. We compare these learning techniques in terms of EDP prediction accuracy, and hardware implementation cost. As mentioned in previous section, in this work we focus on analyzing two heterogeneous CCAs consisting of 8base/4comp, and 4base/2comp cores. In order to perform a comprehensive EDP characterization of studied architectures, we classified all possible configurations (core types and number of threads) into four classes. The first two are *Fully-Base* and *Fully-Composed* configurations that are referred to cases in which the best energy-efficiency is achieved with full utilization of the base and composed core, respectively. In other words, the optimum number of threads is equal to the maximum number of existing base/composed cores. On the other hand, we use *Partially-Base* and *Partially-Composed* configurations when the best number of threads is lower than maximum available cores.

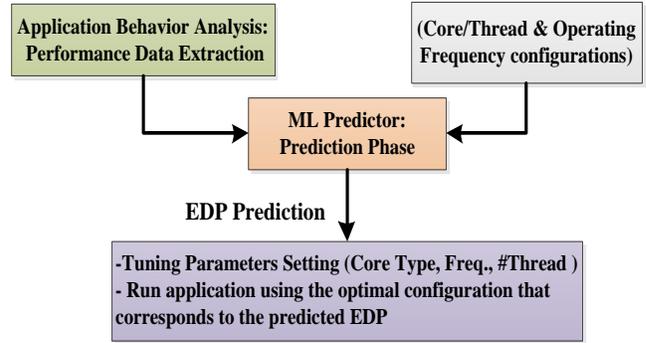

**Figure 5: Energy-efficiency prediction and tuning parameters configuration**

**Optimal Configurations.** Fig. 6 shows the distribution of the optimal configurations for two studied composite core architectures. It demonstrates how the distribution of optimal configurations changes across all studied parallel regions (for all studied applications). In both studied architectures, Fully-Base configuration running at a medium frequency of 2.4 GHz yields the lowest EDP for a majority of studied cases. However, composing cores yields the lowest EDP also for a noticeable number of cases; 37.5% in 8Base/4Comp and 12.5% in 4Base/2Comp. From Fig. 6(a), we observe that overall 62.5% of studied cases benefit from Fully-Base configuration which yields the lowest EDP. Moreover, as shown in Fig. 6(b), for all studied cases in 4Base/2Comp architecture, no Partially-Composed or Partially-Base configurations was selected as the optimum configuration; i.e. in this architecture for maximum energy-efficiency, all cores, either base or composed, need to be allocated to the running multithreaded application. Also, the optimal clock frequency found to be lower than the maximum frequency for the majority of studied cases in both architectures (more than 87.5%). This diagram shows the need to adapt the microarchitecture and application settings to different multithreaded applications for energy-efficiency optimization. Overall, the results confirm a large disparity in the optimum configuration across a large range of tuning parameters, highlighting the importance of developing a predictive method. Also, as the number of base cores in the studied architecture increases, the importance of composing cores to make a larger core is highlighted; more than 32% of studied cases in 8Base/4Comp vs. 12.5% in 4Base/2Comp are corresponding to composite cores.

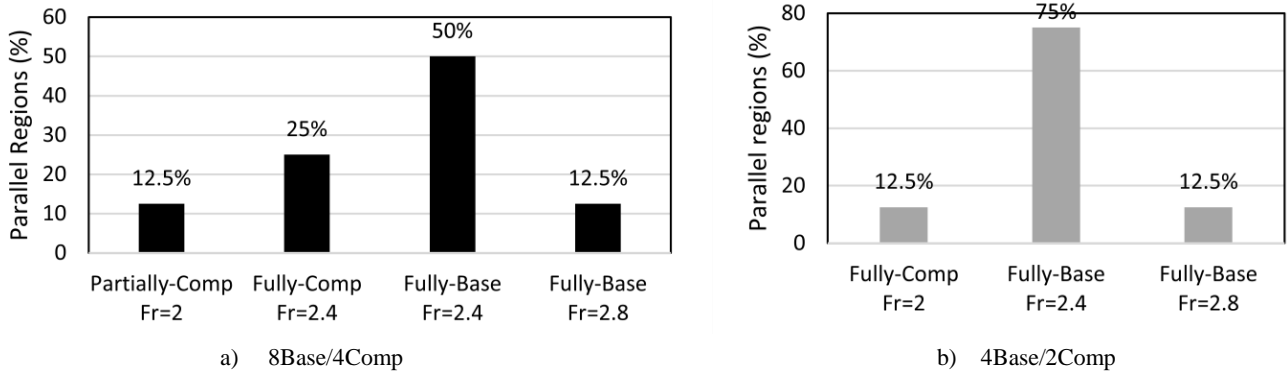

**Fig. 6: The distribution of the optimal configurations for EDP across two different architectures.**

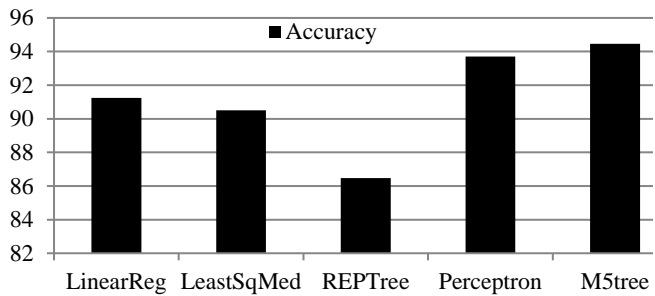

**Fig. 7. EDP accuracy comparison of ML predictors**

**Prediction Accuracy.** Here, we evaluate the accuracy and error rate of our ML-based approach. Error rate and accuracy are important metrics for evaluating the algorithmic prediction and optimization techniques for analyzing and improving any target value in computing systems [12, 31, 45]. In order to evaluate the accuracy of our prediction model, we calculate the value of relative mean absolute error defined as $\frac{|estimated\ value - actual\ value|}{actual\ value} \times 100\%$. This metric indicates the relative difference between the predicted and observed maximum energy-efficiency (EDP) [3, 4]. Fig. 7 shows the accuracy comparison of the machine learning classifiers used for predicting the EDP. As shown, M5Tree achieves close to 94.5% accuracy and outperforms all other classifiers in predicting the energy-efficiency. This tree-based classifier generates a decision list for regression problems using separate-and-conquer process which results in highest EDP accuracy. Next are Perceptron, LinearReg and LeastSqMed predictors, respectively. We implemented a Multi-Layer Perceptron neural network with three layers which is capable of numerical predictions, since neurons are isolated and region approximations can be adjusted independently to each other. Finally, REPTree classifier shows the lowest accuracy as compared to other learning models. REPTree is another fast decision tree learning model, which builds a decision tree using information gain and variance. This model only sorts values for numeric attributes once and missing values are dealt with by splitting the corresponding instances into pieces which negatively impacts the accuracy of this predictor in our EDP prediction problem as compared to other models.

**Table 5. Hardware synthesis comparison of ML predictors**

| ML Algorithm | Latency (cycles@10ns) | Power (W) | Area (LUTs+FFs+DSPs) |
|---|---|---|---|
| LinearReg | 36 | 0.253 | 3071 |
| LeastSqMed | 46 | 0.267 | 3127 |
| MultiLayerPercep | 116 | 0.52 | 20955 |
| M5Tree | 51 | 0.287 | 11120 |
| REPTree | 9 | 0.241 | 2532 |

**Hardware Implementation.** In this section, we discuss the hardware implementation of the machine learning classifiers. We use Vivado HLS compiler to develop the HDL implementation of the classifiers. When it comes to choosing machine learning classifiers for hardware implementation, accuracy of any algorithm is not the only parameter for decision-making [43, 47, 48]. Area, power and latency overhead of ML classifiers are also key factors in selecting a cost-efficient machine learning classifier [30, 49]. While complex algorithms such as neural networks can deliver high accuracy, they will also add significant overhead in terms of hardware implementation. Also, given their complexity, they might be slow in finding the right configuration for scheduling. We are interested in analyzing these overheads when implementing these machine-learning algorithms. The ML algorithm with high accuracy, low area, low power consumption, and low latency is the ideal choice for EDP prediction to guide the scheduling.

The latency, power and area results for implemented machine learning algorithms are shown in Table 5. As can be seen, Multi-Layer Perceptron algorithm results in significant area and latency overhead compare to other learning methods. REPtree decision tree is the fastest algorithm compared to others but comes with the lowest accuracy. M5Tree, another decision tree learning predictor, is the most accurate predictor however it comes with significant area overhead. Clearly the results show some trade-off between accuracy, latency, and area overhead. Therefore, it is important to compare classifiers by taking these parameters into account.

The metric of accuracy over area is a fair ratio to compare the studied predictors. This metric essentially indicates which learning algorithm is the most accurate per unit of silicon area.

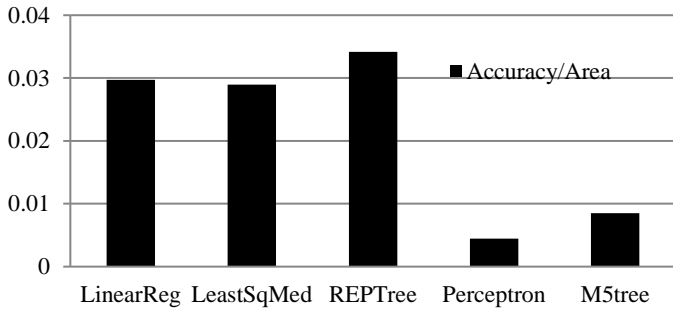

**Fig. 8. Accuracy/Area ratio comparison between ML predictors**

We have shown results of Accuracy/Area in Fig. 8. As can be seen in this figure, REPTree, LinearReg and LeastSqMed classifiers are performing significantly better in terms of accuracy per area compared to highly accurate but complex MultiLayerPerceptron and M5Tree. In addition, if delay is a constraint, REPTree, LinearReg and LeastSqMed classifiers outperform the more complex MultiLayerPerceptron and M5Tree in terms of latency. This is mainly because of REPTree model which doesn't involve complex floating-point operations unlike others and instead, it involves various conditional evaluations. This helps REPTree to achieve lower power consumption as well. However, this is not the case for every tree-based classifier. M5tree which is also a tree-based classifier with higher power consumption has floating point operations. Out of all classifiers, Perceptron performs worst in terms of power consumption and latency mostly because of complex sigmoid function calculations. Comparing based on accuracy/area ratio, the results show REPTree outperforming all other learning algorithms.

## 7. CONCLUSION

Heterogeneous composite core architectures are complex processors with various tuning optimization knobs for improving performance and energy-efficiency. Scheduling multithreaded applications in these architectures is a challenging problem, given various optimization parameters such as application (number of running threads), system (operating voltage and frequency), and architecture (core type, namely big vs. little). In particular, the interplay among these tuning parameters and their influence on energy-efficiency, make the scheduling and tuning even a more challenging problem. In this paper, we respond to this challenge by developing a scheduling and tuning solution for this class of architecture. The space for tuning configuration parameters in a composite core architecture is large, and our analysis indicate that there is no unique solution for the most energy-efficient configuration for different multithreaded applications, calling for developing a model to predict energy-efficiency for various tuning parameters. In response, we present a systematic approach for energy-efficiency prediction using various machine learning algorithms. We develop five machine learning-based models for estimating energy-efficiency of multithreaded applications in composite core architecture. Our proposed ML-based model takes hardware performance counters information at run-time from a multithreaded application, and it predicts the most energy-efficient configurations based on run-time analysis and sets the number of threads and operating frequency. It also decides whether to compose little cores into big cores. The results show significant energy-efficiency prediction accuracy of as high as 94% across studied applications. We compared these algorithms in terms of their accuracy, latency, power and area overhead. Our results show that although using non-linear regression or neural network models such as M5Tree or MultiLayerPerceptron provides more accurate energy-efficiency prediction compared to simple regression or decision tree-based models, they significantly increase complexity of the design in terms of power and area overhead. They are also noticeably slower than regression-based or decision tree models.